\documentclass[12pt]{article}
\usepackage{graphicx}
\usepackage{capt-of}
\usepackage{caption}

\newcommand\pubnumber{DPF2015-377}
\newcommand\pubdate{\today}

\def\napoli{National Centre for Nuclear Research\\
Warsaw, POLAND}
\def\support{\footnote{Work supported by European Union grant H2020-MSCA-RISE-2014 641540.}}

\def\Title#1{\begin{center} {\Large #1 } \end{center}}
\def\Author#1{\begin{center}{ \sc #1} \end{center}}
\def\Address#1{\begin{center}{ \it #1} \end{center}}

\newcommand\pubblock{\rightline{\begin{tabular}{l} \pubnumber\\
         \pubdate  \end{tabular}}}
\newenvironment{Abstract}{\begin{quotation}  }{\end{quotation}}
\newenvironment{Presented}{\begin{quotation} \begin{center} 
             PRESENTED AT\end{center}\bigskip 
      \begin{center}\begin{large}}{\end{large}\end{center} \end{quotation}}

\def\beq{\begin{equation}}
\def\eeq#1{\label{#1}\end{equation}}
\def\eeqn{\end{equation}}

\def\beqa{\begin{eqnarray}}
\def\eeqa#1{\label{#1}\end{eqnarray}}
\def\eeqan{\end{eqnarray}}

\let\bar=\overbar

\def\Dslash{\not{\hbox{\kern-4pt $D$}}}
\def\dslash{\not{\hbox{\kern-2pt $\del$}}}

\def\msb{{\bar{\ssstyle M \kern -1pt S}}}

\begin{document}
\begin{titlepage}
\pubblock

\vfill
\Title{Searching for Dark Matter Annihilation into Neutrinos with Super-Kamiokande}
\vfill
\Author{Katarzyna Frankiewicz\support}
\Address{\napoli}
\vfill
\begin{Abstract}
This work presents indirect searches for dark matter (DM) as WIMPs (Weakly Interacting Massive Particles)
 using neutrino data recorded by the Super-Kamiokande detector from 1996 to 2014.
 The results of the search for WIMP-induced neutrinos from the Sun and the Milky Way are discussed. 
 We looked for an excess of neutrinos from the Sun/Milky Way direction compared to the expected atmospheric neutrino background. 
 Event samples including both electron and muon neutrinos covering a wide range of neutrino energies (GeV to TeV) were used,
 with sensitivity to WIMP masses down to tens of GeV. Various WIMP annihilation modes were taken into account in the analyses.
\end{Abstract}
\vfill
\begin{Presented}
DPF 2015\\
The Meeting of the American Physical Society\\
Division of Particles and Fields\\
Ann Arbor, Michigan, August 4--8, 2015\\
\end{Presented}
\vfill
\end{titlepage}
\def\thefootnote{\fnsymbol{footnote}}
\setcounter{footnote}{0}
\section{Dark matter detection}
Dark matter composes about 27\% of the total mass-energy of the Universe~\cite{planck}. Detection and elucidating of its nature is one of the main goals of astrophysics and particle physics nowadays.
Observation of WIMPs present in the Galactic halo may be attempted directly
via elastic scattering off nuclei in the detectors or indirectly through detection of the products of their annihilations, such as charged particles, photons or neutrinos. The latter, can be produced directly or in subsequent decays of mesons and leptons. 
Neutrinos can provide very good information on their source position while traversing unaffected through galactic scales.
Moreover, their energy remain unchanged during propagation providing valuable information about energy spectra generated in DM annihilation processes. 
\section{Super-Kamiokande detector and atmospheric $\nu$'s}
Super-Kamiokande (SK) is the 50 kton water Cherenkov detector located in the Kamioka Observatory of the Institute for Cosmic Ray Research, 
University of Tokyo~\cite{sk_det}. The observatory was designed to search for a proton decay, study solar, atmospheric and man-made neutrinos, 
and keep watch for supernovae. Detection of neutrino interactions is based on observation of charged particles, primarly leptons, 
which may produce Cherenkov radiation while moving faster than $c$ in water. The Cherenkov light projected onto the walls of the detector and recorded by photomultiplier detectors, allows to reconstruct energy, direction and flavour of produced lepton.
\newline
\newline 
Among all neutrino sources it is worth to discuss atmospheric neutrinos as they have the same energy range as it is expected for DM induced neutrinos. Atmospheric neutrinos are produced in interactions of cosmic rays with atomic nuclei in the Earth's atmosphere. 
Their average energy is of several hundreds of MeV and the energy spectra has a long high energy tail reaching TeV scale. 
If the charged lepton produced in neutrino interaction stops in the inner detector, this type of event is classified as fully-contained (FC). 
If a high energy lepton exits the inner detector and deposits energy also in the outer veto region, event
is classified as partially-contained (PC). The energies of PC events are typically 10 times higher than those producing FC events.
Neutrinos also interact with the rock surrounding the detector and may produce high energy muons which intersect the tank. Downward-going muons produced in interactions of neutrinos cannot be distinguished from the constant flux of cosmic ray muons. However, muons travelling
in upward direction (UPMU) must be neutrino induced. 
The energies of the neutrinos
which produce stopping muons are roughly the same as for
PC events, $\sim$10 GeV. Upward through-going
events are significantly more energetic, the parent
neutrino energy for these events is about 100 GeV on average.
\newline
\newline
The discussed search for DM-induced $\nu$'s from the Sun is based on atmospheric neutrino data collected with SK detector in years 1996-2012 (SK-I, SK-II, SK-III and SK-IV data taking periods), corresponding in total to 3902.7 livetime-days for FC/PC and 4206.7 livetime-days for UPMU events.
The second analysis regarding Galactic Halo as a source is extended with data collected in 2013 and 2014, which in total corresponds to 4223.3 livetime-days for FC/PC and 4527.0 livetime-days for UPMU events.
\section{Analysis}
In the conducted searches it is assumed that atmospheric neutrino data collected with the Super-Kamiokande detector could
be described by two components: WIMP-induced neutrinos (signal) and atmospheric neutrinos (background).
The best combination of signal and background that would fully explain the data is tried to find using a fit
method. Both, the signal and background prediction are based on the Monte Carlo (MC) simulations. The signal
contribution can be govern by the normalization parameter and varied in a fit. The background normalization
and shape can be fitted throughout the values of the atmospheric neutrino oscillation parameters and through the
values of systematic uncertainty terms - some of them govern the absolute normalization of atmospheric
neutrino fluxes.
\newline
\newline
The background simulation is available for the atmospheric neutrinos in large datasets corresponding to
around 2000 years of the running of the experiment. Oscillations of neutrinos are taken into account in those
predictions. The simulations assume realistic fluxes of neutrinos, their interaction in the detector or in the
surrounding rock. They include the detector response for the produced Cherenkov light. The same set of reduction and classification cuts is applied to simulated events as to the real data. 
\newline
\newline
In order to simulate the signal neutrinos from DM annihilation in the Milky Way and in the Sun core, DarkSUSY [4] and WimpSim [5] are used. Given the expected signal characteristics at the production point, neutrino propagation under the assumption of three-favour oscillations is applied. As a next step, neutrino interactions in the detector along with the detector response are obtained for each event from the MC set. Therefore, the final signal simulation set reflects the characteristics for expected WIMP-induced $\nu$'s: reconstructed angular distributions of neutrino directions are peaked from the direction of the source and their energy spectra expected for various WIMP annihilation channels are reproduced and take into account detector effects. Each analysis is performed in the coordinate system in which the expected signal is can be distinguished the most effectively from the atmospheric neutrino background.
\newline
\newline
Separate MC sets are used to simulate signal and background to avoid correlations. The models are prepared for various masses of the relic particles and different DM annihilation scenarios. 
There are 18 data samples used in the analyses, including both e-like and $\mu$-like event categories. Each sample is binned in momentum and the cosine of the angle between event direction and direction of the Sun ($\cos\theta_{SUN}$) or the Galactic Center ($\cos\theta_{GC}$).
The signal and background contribution to the analysis bins
differs and enables for an effective discrimination between those two sources of neutrinos. The signal is largely peaked in the direction of the Sun or Galactic Center, while the background is
not (see Fig.~\ref{sub_ex}). Additional constraint can be obtained based on neutrino energy information and proportions of
signal/background in various event subsamples.
Based on the developed simulations one could attempt to perform a fit to the collected data and estimate how
many DM-induced $\nu$'s can be contained in SK data so far. The fitting procedure is repeated
for various masses of WIMP particles and various annihilation channels. 
\section{Results}
\subsection{Solar WIMP search}
\begin{figure}[b!]
  \begin{minipage}[t]{0.5\textwidth}
\captionsetup{width=.95\linewidth}
\centering
    \includegraphics[width=0.95\textwidth]{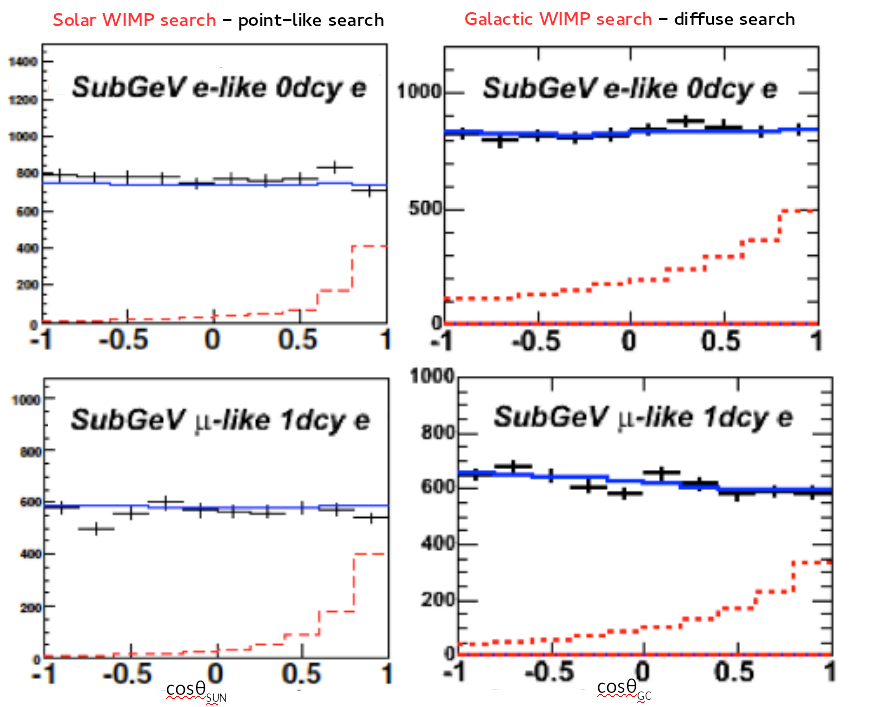}
\caption{\scriptsize Angular distributions of SK I-IV data (black crosses), atm $\nu$ background MC (normalized to data live time, solid blue) and WIMP neutrino signal MC (dashed red) for 10 GeV WIMPs annihilating into $b{\bar b}$ (example only for 2 data samples out of 18).}
\label{sub_ex} 
 \end{minipage}\hfill
  \begin{minipage}[t]{0.5\textwidth}
\captionsetup{width=.95\linewidth}
\centering
    \includegraphics[width=0.9\textwidth]{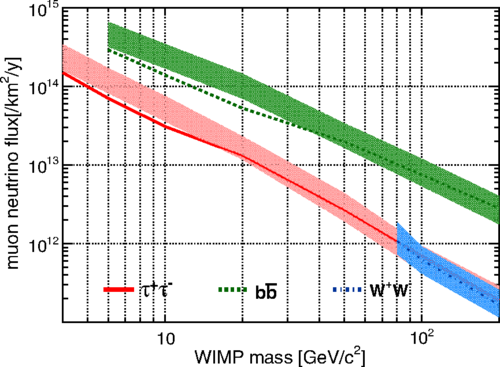}
    \caption{\scriptsize The 90\% CL upper limit on total integrated $\nu_{\mu}$ flux from WIMP annihilation in the Sun for $\tau^{+}\tau^{-}$ (solid red), $b\bar{b}$ (dashed green) and $W^{+}W^{-}$ (dot-dashed blue) annihilation channels. The shadowed regions show 1$\sigma$ bands of the sensitivity study results.}
     \label{fig:flux_lim}
  \end{minipage}
\end{figure}
\begin{figure}[ht!]
	\begin{minipage}[t]{0.5\textwidth}
\captionsetup{width=.95\linewidth}
	\centering
	\includegraphics[width=.9\textwidth,angle=0]{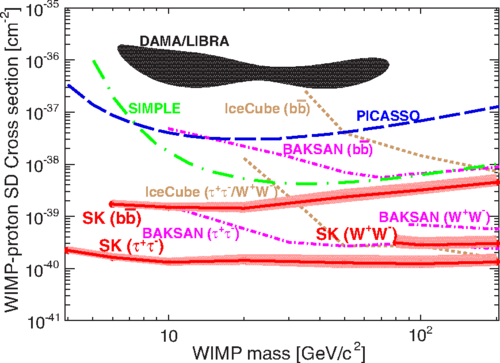}
	\caption{\scriptsize 90\% CL upper limits on SD WIMP-proton cross section calculated at DarkSUSY~\cite{darksusy} default are shown in red
solid with uncertainty bands to take account uncertainties in the capture rate for the $b\bar{b}$, $W^{+}W^{-}$ and $\tau^{+}\tau^{-}$ channels from
top to the bottom. Also limits from other experiments: IceCube~\cite{icecube} (dashed brown), BAKSAN~\cite{baskan} (dot-dashed pink), PICASSO~\cite{picasso} (long-dashed blue) and SIMPLE~\cite{simple} (long dot-dashed green) are shown. The black shaded region is the 3$\sigma$ CL signal claimed by DAMA/LIBRA~\cite{dama}.}
	\label{SD}
	\end{minipage}
	\begin{minipage}[t]{0.5\textwidth}
\captionsetup{width=.95\linewidth}
	\centering
	\includegraphics[width=.9\textwidth,angle=0]{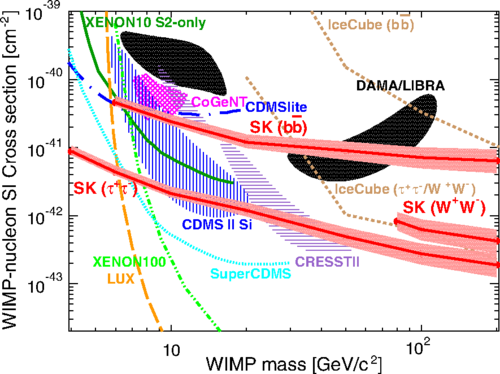}
	\captionof{figure}{\scriptsize 90\% CL upper limits on the SI WIMP-nucleon cross section. Also event excesses or annual modulation signals reported by other experiments: DAMA/LIBRA~\cite{dama} (black regions, 3$\sigma$ CL), CoGeNT~\cite{cogent} (magenta diagonally cross-hatched region, 90\% CL), CRESSTII~\cite{cresst} (violet horizontally-shaded regions, 2$\sigma$ CL), CDMS II Si~\cite{cdms} (blue vertically-shaded region, 90\% CL), and limits: IceCube~\cite{icecube} (dashed brown), SuperCDMS~\cite{supercdms} (dotted cyan), CDMSlite~\cite{cdmslite} (long dot-dashed blue), XENON10 S2-only~\cite{xenon} (dash triple dot dark green), XENON100~\cite{xenon100} (dash double dot green) and LUX~\cite{lux} (long-dashed orange) are shown.}
	\label{SI}
	\end{minipage}
	\end{figure}

For all tested WIMP hypotheses, no contribution of DM-induced $\nu$'s has been found. 90\% CL upper limits were set for the tested WIMP masses and annihilation channels assuming that the obtained $\chi^{2}$ values approximately follow a normal distribution. The derived 90\% upper limit on the muon-neutrino flux from DM annihilations in the Sun in shown in Fig.~\ref{fig:flux_lim}. This limit was converted into the upper limit on WIMP-nucleon cross-section using DarkSUSY 5.0.6~\cite{darksusy}. Only a single type of WIMP interaction with a nucleus, either an axial vector interaction in which WIMPs couple to the nuclear spin (spin dependent, SD) or a scalar interaction in which WIMPs couple to the nucleus mass (spin independent, SI) was assumed. Standard DM halo with local density 0.3 GeV/cm${}^3$~\cite{halo1}\cite{halo2}, a Maxwellian velocity distribution with an RMS velocity of 270 km/s and a solar rotation
speed of 220 km/s were considered. The results are plotted together with other experimental results
in Fig.~\ref{SD} for SD coupling and Fig.~\ref{SI} for SI coupling for the isospin-invariant case.
The uncertainties related to the WIMP capture process are indicated by the shadowed regions (detailed description can be found in~\cite{koun}.
\subsection{Galactic WIMP search}
\begin{figure}[h]
	\begin{minipage}[t]{0.5\textwidth}
        \captionsetup{width=.95\linewidth}
	\centering
	\includegraphics[width=1.\textwidth,angle=0]{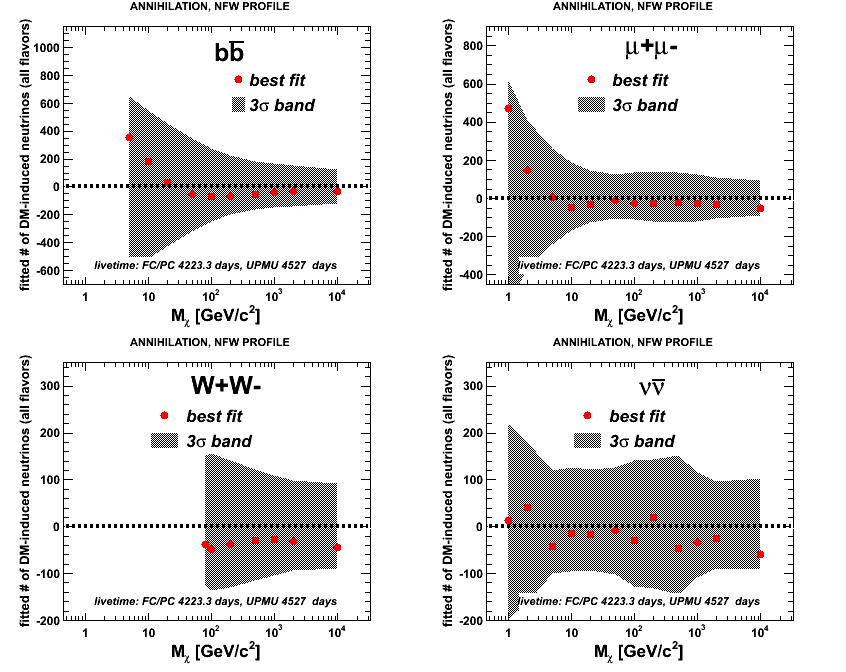}
	\captionof{figure}{\scriptsize Fitted number of WIMP-induced $\nu$'s of all flavors from WIMP annihilation into $b\bar{b}$, $\mu^{+}\mu^{-}$, $W^{+}W^{-}$ and $\nu\bar{\nu}$ as a function of the mass relic particles. The gray band corresponds to 3$\sigma$ sensitivity with null WIMP
	hypothesis.}
	\label{fig:fitGC}
	\end{minipage}
	\begin{minipage}[t]{0.5\textwidth}
     \captionsetup{width=.95\linewidth}
	\centering
	\includegraphics[width=1.\textwidth,angle=0]{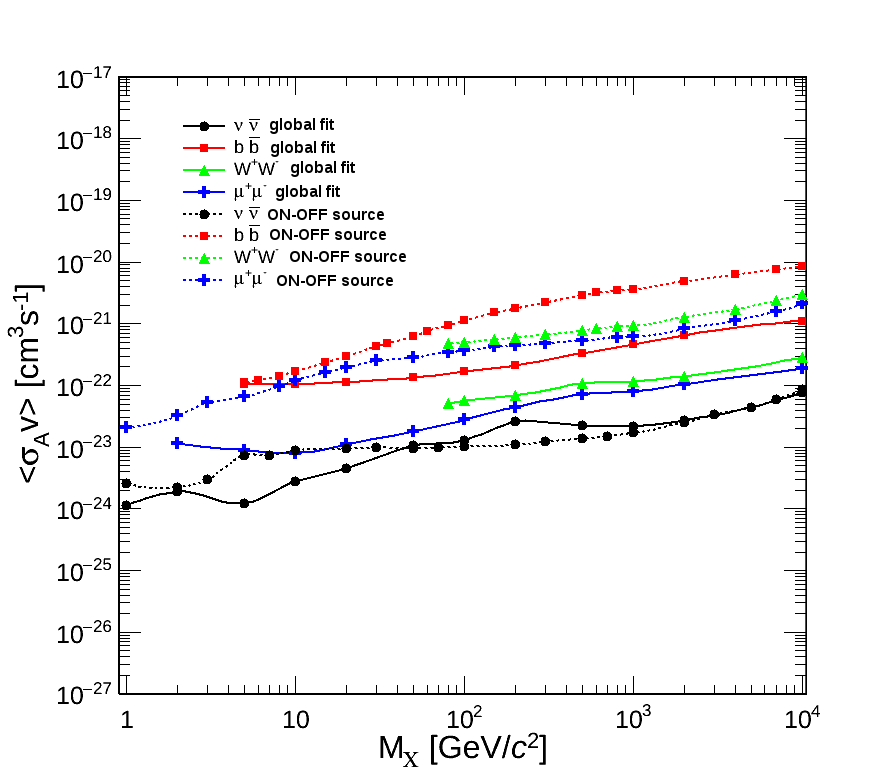}
	\captionof{figure}{\scriptsize 90\% CL upper limits on DM self-annihilation cross
section $\langle\sigma_{A}V\rangle$ (region above the lines is excluded) from global fit (solid) and ON-OFF souce (doted) analyses. Limits for $\nu\nu$ (black circles), $b\bar{b}$ (blue crosses), $W^{+}W^{-}$ (red squares) and $\mu^{+}\mu^{-}$ (green triangles) annihlation modes are based on expected signal intensity (DM squared density) from NFW halo profile~\cite{nfw}.}
\label{fig:limitGC}
	\end{minipage}
	\end{figure}

No significant signal contribution of DM-induced $\nu$'s from the Milky Way is allowed by the data (Fig.~\ref{fig:fitGC}). Points shown in this figure are not independent as the same set of data is used in the fit for every WIMP mass hypothesis. This result can be translated into upper 90\% CL limit on the fitted number of DM-induced $\nu$'s using the bayesian approach~\cite{bayes} which enables to set limit when the result falls into unphysical region (fitted number of WIMPs is negative). Based on the limit on the number of DM-induced $\nu$'s, corresponding limit on DM-induced diffuse neutrino flux can be derived as a function of $M_{\chi}$ and translated into DM self-annihilation
cross section $\langle\sigma_{A}V\rangle$ which is shown in Fig.~\ref{fig:limitGC}. 
Independly, the second analysis based on the concept of ON- and OFF-source regions was conducted. This regions are constructed in a way that the signal is expected only in ON-source region, but the level of background is the same in both regions. The difference in number of neutrino events between ON- and OFF-source regions corresponds to the difference in signal events and is directly proportional to the $\langle\sigma_{A}V\rangle$. In this way the background can be estimated directly from the data. This method is independent on atmospheric MC simulations and related systematic uncertainties as they should equally affect ON- and OFF-source regions.

\end{document}